4

# π-Edge: A Low-Power Edge Computing System for Real-Time Autonomous Driving Services

Jie Tang, Shaoshan Liu, Bo Yu, and Weisong Shi *Fellow, IEEE*

*Abstract* – **To simultaneously enable multiple autonomous driving services on affordable embedded systems, we designed and implemented π-Edge, a complete edge computing framework for autonomous robots and vehicles. The contributions of this paper are three-folds: first, we developed a runtime layer to fully utilize the heterogeneous computing resources of low-power edge computing systems; second, we developed an extremely lightweight operating system to manage multiple autonomous driving services and their communications; third, we developed an edge-cloud coordinator to dynamically offload tasks to the cloud to optimize client system energy consumption. To the best of our knowledge, this is the first complete edge computing system of a production autonomous vehicle. In addition, we successfully implemented π-Edge on a Nvidia Jetson and demonstrated that we could successfully support multiple autonomous driving services with only 11 W of power consumption, and hence proving the effectiveness of the proposed π-Edge system.**

## I. Introduction

Many major autonomous driving companies, such as Waymo, Baidu, and Uber, and several others are engaged in a competition to design and deploy the ultimate ubiquitous autonomous vehicle which can operate reliably and affordably, even in the most extreme environments. Yet, to build such an autonomous vehicle, the cost for all sensors could easily be over $100,000, the cost for the computing system adds another $30,000, resulting in an extremely high cost for each vehicle: for instance, a demo autonomous vehicle can easily cost over $300,000 [1].

Further, beyond the unit cost, it is still unclear how the operational costs for High Definition (HD) map creation and maintenance will be covered. In addition, even with the most advanced sensors, having autonomous vehicles co-exist with human-driven vehicles in complex traffic conditions remains a dicey proposition.

As a result, unless we can significantly drop the costs of sensors, and computing systems as well as dramatically improve localization, perception, and decision making algorithms in the next few years, autonomous driving will not be universally adopted.

Addressing these problems, a reliable autonomous vehicle, the DragonFly Pod (Figure 1), has been developed by us, for a total cost under $10,000 when mass-produced and for low-speed scenarios, such as university campuses, industrial parks, and areas with limited traffic.

The DragonFly pod supports three basic services, real-time localization through Simultaneous Localization And Mapping (SLAM), real-time obstacle detection through computer vision, and speech recognition for user interaction [28].

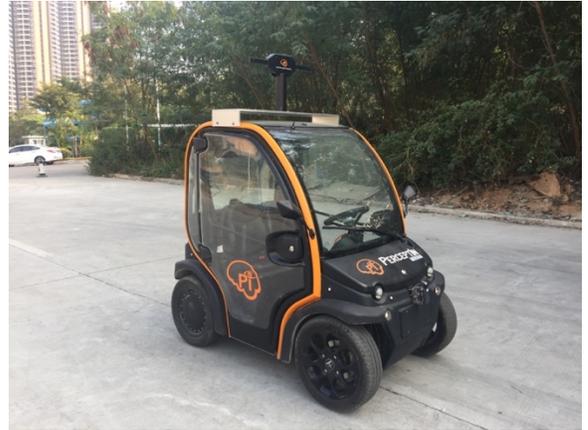

Fig 1. PerceptIn DragonFly Pod

To achieve this, one major challenge is to simultaneously enable localization, object detection, and speech recognition on an affordable low-power edge computing system. For localization, we utilize visual information to provide real-time updates of the vehicle's location; for perception, we utilize visual information to recognize the obstacles blocking the vehicle, such that the vehicle can adjust its action accordingly to avoid the obstacles; for speech recognition, the passengers can give a speech command within the vehicle at any time to interact with the vehicle.

Simultaneously supporting these services on a low-power edge computing system is extremely challenging. First, these services have complex pipelines, are computation-intensive, and often have tight real-time requirement. For example, inertial measure unit (IMU) data can rush in at a rate as high as 1 KHz in SLAM, meaning that the computation pipeline needs to process sensor data at a speed to be able to produce 1,000 position updates in a second; making the longest stage of the pipeline cannot take more than 1 millisecond to process. Moreover, the samples form a time series and are independent. This means that the incoming samples often cannot be processed in parallel. For Convolution Neural Networks (CNN), the camera may capture pictures at a rate of 60 frames

―――――――――――――――――――――
Jie Tang is currently with South China University of Technology (e-mail: cstangjie@scut.edu.cn).
Shaoshan Liu and Bo Yu are currently with PerceptIn (e-mail: shaoshan.liu@perceptin.io, bo.yu@perceptin.io).
Weisong Shi are currently with Wayne State University (e-mail: weisong@wayne.edu)

per second (FPS), meaning that the CNN pipeline needs to be able to extract meaningful features and recognize the objects within 16 ms. Similarly, speech recognition imposes strong real-time requirements in order for the communication to be interactive. Second, the edge computing system has extremely limited energy budget as it runs on the vehicle's battery. Therefore, it is imperative to optimize power consumption in these scenarios.

As far as we know, this is the first paper on a complete edge computing system of a production autonomous vehicle. The contributions of this paper are as follows:

First, to fully utilize the heterogeneous computing resources of low-power edge computing systems, we developed a runtime layer as well as scheduling algorithms to map autonomous driving computing tasks to heterogeneous computing units to achieve optimal performance and energy efficiency. More details of the runtime layer design can be found in Section IV.

Second, existing operating systems for robotic workloads, such as Robot Operating Systems (ROS) often impose very high communication and computing overheads, and thus not suitable for systems with limited computing resources and energy consumption constraints. To manage multiple autonomous driving services and their communications, we developed an extremely lightweight operating system, π-OS, for low-power edge computing systems, and proved the effectiveness of π-OS. More details of π-OS can be found in Section V.

Third, at times offloading computing from edge client to cloud leads to energy efficiency, but whether to offload and how to offload remains an unsolved problem. To address this problem, we developed an edge-client-cloud coordinator to dynamically offload tasks to the cloud to optimize edge computing system energy consumption. More details of the offloading algorithms can be found in Section VI.

Last but not least, we successfully integrated these components into our proposed π-Edge system and implemented it on Nvidia Jetson. We demonstrated that we could successfully support multiple autonomous driving services with only 11 W of power consumption, and thus proving the effectiveness of the proposed π-Edge system. More details on the system integration can be found in Section VII.

## II. AUTONOMOUS DRIVING SERVICES

Before going into the details of the π-Edge system design, let us briefly examine the services needed in autonomous vehicle systems. As shown in Figure 2, a fully functioning autonomous vehicle must be able to perceive its environment and safely navigate on the basis of multiple sensors rather than a human driver [1, 2]. These sensors typically include laser imaging detection and ranging (LiDAR), a global positioning system (GPS), an inertial measurement unit (IMU), various cameras, or any combination of these sensors.

Each vehicle uses sensor inputs in localization—the process of understanding its environment—and in making real-time decisions about how to navigate within that perceived environment. These tasks involve processing a high volume of sensor data and require a complex computational pipeline. For instance, in existing designs, an autonomous car is typically equipped with multiple computing servers, each with several high-end CPUs and GPUs. Consequently, autonomous vehicles computing systems usually impose very high power consumption — often thousands of watts [20].

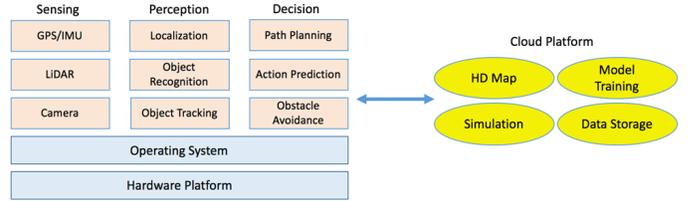

Fig 2. Autonomous Driving Technology Stack

In this paper we focus on the services of the DragonFly pod, a low-speed autonomous vehicle described in the introduction. The core technologies enabling autonomous driving are simultaneous localization and mapping (SLAM) for localization, convolutional neural networks (CNN) for object recognition, and speech recognition for user interactions.

SLAM refers to the process of constructing or updating the map of an unknown environment while simultaneously keeping track of the location of the agent [3]. It is a complex pipeline that consists of many computation-intensive stages, each performing a unique task. CNN-based object recognition consists of multiple processing layers to model high-level abstractions in the data [4]. In recent years, CNN has outperformed many traditional methods thanks to the significant improvements from the computer vision field. Speech recognition can utilize methods such as Gaussian mixture model (GMM) [5] or deep neural network (DNN) [6] for speech frame classification; then these two approaches can be used together with the hidden Markov model (HMM) and Viterbi algorithm to decode frame sequences.

### A. Localization

Figure 3 shows a simplified version of the general SLAM pipeline which operates as follows:

1.) The Inertial Measurement Unit, or IMU, consists of a 3 Degrees of Freedom (DoF) gyroscope to measure angular velocity and 3-DoF accelerometers to measure acceleration along the three axes. The 6-DoF IMU produces data points at a high rate, which is fed to the propagation stage.

2.) The main task of the Propagation Unit is to integrate the IMU data points and produce a new position. Since IMU data is received at a fixed interval, by integrating the accelerations twice over time, we can derive the displacement of the agent during the last interval. However, since the IMU hardware usually has bias and inaccuracies, we cannot fully rely on propagation data, lest the positions produced gradually drift from the actual path.

3.) To correct the drift problem, we use a camera to capture frames along the path at a fixed rate, usually at 60 FPS.

4.) The frames captured by the camera can be fed to the Feature Extraction Unit, which extracts useful corner features and generates a descriptor for each feature.

5.) The features extracted can then be fed to the Mapping Unit to extend the map as the agent explores. Note that by map we mean a collection of 3D points in space, where each 3D point would correspond to one or more feature points detected in the Feature Extraction Unit.

6.) The features detected would also be sent to the Update Unit which compares the features to the map. If the detected features already exist in the map, the Update Unit can then derive the agent's current position from the known map points. By using this new position, the Update Unit can correct the drift introduced by the Propagation Unit. The Update Unit updates the map with the newly detected feature points as well.

In this implementation, we use our proprietary SLAM system [7, 8] that utilizes a stereo camera for image generation at 60 FPS, with each frame having the size of 640 X 480 pixels. In addition, the IMU device generates 200 Hz of IMU updates (three axes of angular velocity and three axes of acceleration).

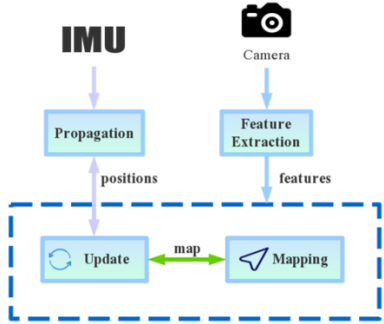

Fig 3. Visual Inertial SLAM Execution

### B. Object Recognition

Convolutional Neural Network (CNN) is a type of Deep Neural Network that is widely used in object recognition tasks [4]. Figure 4 shows a simplified version of the general CNN evaluation pipeline, which usually consists of the following layers:

1.) The Convolution Layer contains different filters to extract different features from the input image. Each filter contains a set of "learnable" parameters that will be derived after the training stage.

2.) The Activation Layer decides whether to activate the target neuron or not. Common activation functions include the saturating hyperbolic tangent function, the sigmoid function, and the rectified linear units.

3.) The Pooling Layer reduces the spatial size of the representation to reduce the number of parameters and consequently the computation in the network.

4.) The Fully Connected Layer is where neurons have full connections to all activations in the previous layer. It derives the labels associated with the input data.

Note that in a normal network we would have multiple copies of the convolution, activation, and pooling layers. This way, the network can first extract low-level features, and from the low-level features it derives high-level features, and at the end it reaches the Fully Connected Layer to generate the labels associated with the input image.

In this implementation, we use the Single Shot Multi-Box Detector [9], which discretizes the output space of bounding boxes into a set of default boxes over different aspect ratios and scales per feature map location.

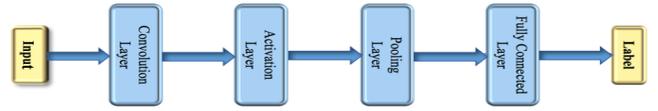

Fig 4. Simplified CNN Inference Engine

### C. Speech Recognition

A generic pipeline of speech recognition is shown in Figure 5, which can be divided into the following stages:

1.) First the speech signal goes through the feature extraction stage, which extracts feature vector. This is where we utilize a GMM-based feature extractor.

2.) Then the extracted feature vector is fed to the decoder, which takes an acoustic model, a pronunciation dictionary, and a language model as input. The decoder then decodes the feature vector into a list of words.

Note that in this implementation we utilize the speech model presented in [10], which uses GMM for classification and HMM for decoding.

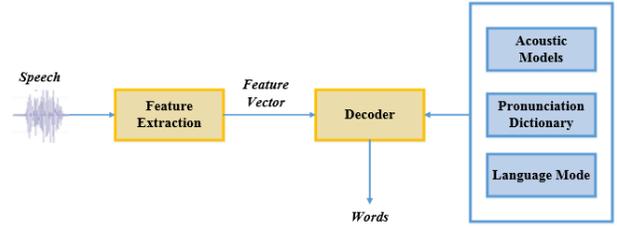

Fig 5. Speech Recognition Engine

## III. π-EDGE ARCHITECTURE

As discussed in the introduction, to enable the affordable and reliable DragonFly Pod, we need to integrate multiple autonomous driving services onto a low-power edge computing device.

This poses several challenges: first, as the edge devices usually consist of heterogeneous computing units, computing and energy efficiency can only be achieved if the heterogeneous computing resources can be fully utilized. However, mapping different tasks dynamically to different computing units is complex and challenging, and we do not want to expose this complexity to the service developer. Hence, we need a runtime to dynamically manage the underlying heterogeneous computing resources as well as schedule different tasks onto these units to achieve optimal performance or energy efficiency.

Second, with multiple services running on a resource-constrained device, we need an extremely lightweight operating system to manage these services and facilitate the communications between them. Existing operating systems, such as ROS, impose very high computing and memory overheads and thus not suitable for our design.

Third, one way to optimize energy efficiency and to improve the computing capability of edge client devices is to offload computing workloads to the cloud when possible. However, dynamically deciding whether to offload, and how to offload is another complex and challenging task. To achieve, we need to develop algorithms to handle offloading.

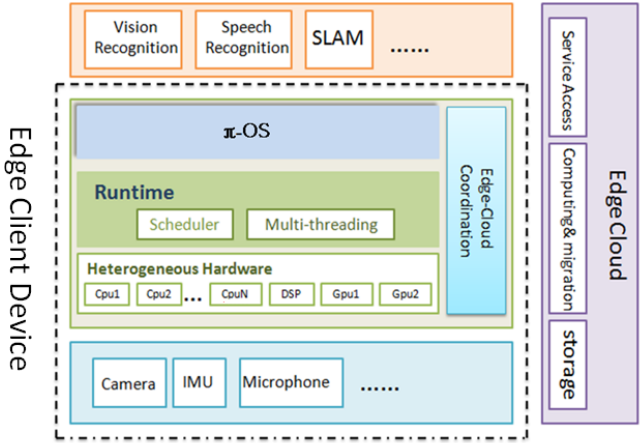

Fig 6. π-Edge Architecture

As shown in Figure 6, π-Edge is designed to address these exact problems. At the application layer, currently π-Edge supports localization, obstacle detection, and speech recognition. Then to integrate these services we developed π-OS, an extremely lightweight operating system that manages various services and facilitates their communications with almost zero overheads. π-OS serves as the basic communication backbone. Comparing to ROS, π-OS is extremely lightweight and optimized for both inter-process communications on the same device, as well as inter-device communications.

Below π-OS is the runtime layer, which implements two functions: first, it provides an abstraction of the underlying heterogeneous computing resources through and provides acceleration operations; second, it implements a two-tier scheduling algorithm to manage the mapping of tasks on heterogeneous hardware systems.

In addition, in order to effectively control the energy consumption of autonomous vehicles, π-Edge contains an edge-cloud coordinator to dynamically offload some tasks to the cloud to achieve optimal energy efficiency. Specifically, taking into account the mobility of vehicles and the cloud availability, we developed an algorithm to dynamically determine the weight of task offload as well as cloud service node selection. We delve into each of these components in the next few sections.

## IV. THE RUNTIME LAYER

The first major contributions of this paper is the design and implementation of the runtime layer to dynamically map various tasks onto the underlying heterogeneous computing units. This runtime layer is crucial to simultaneously enable multiple autonomous driving tasks on computing and energy resource constrained edge computing systems.

Figure 7 shows the design of the π-Edge runtime. To manage the underlying heterogeneous computing resources, we utilized OpenCL, an open standard for cross-platform, parallel programming of diverse computing units [31]. OpenCL provides the interface for π-Edge to schedule and dispatch various applications to the underlying heterogeneous computing resources.

On top of OpenCL, we designed and implemented a scheduler to manage and dynamically dispatch incoming tasks. Our scheduler is a two-layer design: the inter-core scheduler dispatches incoming tasks, such that they are added to the queue of different heterogeneous computing units based on resource availabilities and task characteristics.

Then within each computing resource queue, the inner-core scheduler determines the order of execution based on the priority and characteristics of each task.

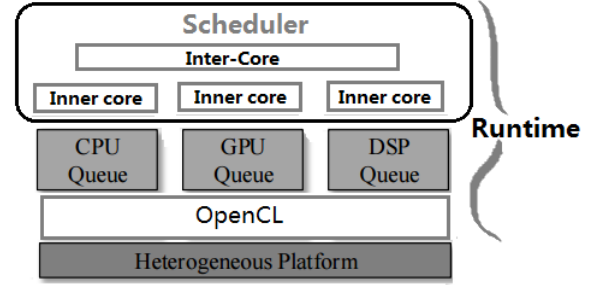

Fig 7. the runtime layer design

### A. Inter-core scheduler

In detail, when triggered, the inter-core scheduler first examines the loads of all computing units, and schedules tasks to the processor with minimal execution time to balance the load of the processor. Note that we constrain that each task is only scheduled to one type of computing unit and thus there is no dependencies between tasks. The independent task scheduling method in heterogeneous multi-core processor is a NP problem, should follow two principles in [36]: 1.) Matching: each task should be mapped to the processor that minimizes its execution time so that the task can be completed as quickly as possible. 2.) Load balancing: there should be no overload or underuse for each processor. In order to better describe the Inter-Core scheduling algorithm, let us first define some basic concepts as follows [37]:

*Metatask*: a collection of independent tasks to be assigned
$$Metatask = \{t_i | 0 < i < \gamma\}$$
*U*: the task queue accommodating all unmapped tasks. When scheduling starts, $U = Metatask$.
*Q*: the set of heterogeneous cores for scheduling
$$Q = \{c_j | 0 < j < \chi\}$$
*ETC($t_i$, $c_j$)*: We use the ETC-table to store the ETC value.

Each *ETC($t_i$, $c_j$)* in table shows the *Expected execution time* of task $t_i | t_i \in Metatask$ when it is running on core $c_j | c_j \in Q$. It includes the computation time as well as the time to move the executable and data associated with task $t_i$ from their known source to the core. For cases when it is impossible to execute task $t_i$ on core $c_j$, the value of *ETC($t_i$, $c_j$)* is set to infinity.

*mat($c_j$)*: Machine availability time for core $c_j$, $0 < j < \chi$. It is the earliest time core $c_j$ can complete the execution of all the tasks that have previously been assigned to it.
*ct($t_i$, $c_j$)*: The completion time for a new task $t_i$ on core $c_j$.
$$ct(t_i, c_j) = mat(c_j) + ETC(t_i, c_j), t_i \in U$$
*Makespan*: The execution time of Metatask.
$$Makespan = \max ct(t_i, c_j), t_i \in Metatask, c_j \in Q$$

The problem of heterogeneous multi-core scheduling can be defined to find out a list of dispatch order can guarantee minimal makespan.

Task scheduling in heterogeneous environments is a NP problem. Therefore, it is very difficult to reach the best solution in one algorithm. In [34], authors made a comparison of Min-Min [29], Max-Min [39], Genetic Algorithm (GA) [40] and other 11 independent task scheduling algorithms. The results show that Min-Min has the best comprehensive performance, thus we use Min-min as the baseline for comparing in our study.

Min_min heuristic begins with the set $U = Metatask$. Then, the set of minimum completion times, M=[min $ETC(t_i, c_j)$| $t_i \in U$, $c_j \in Q$.], is found. Next, the task with the overall minimum completion time from M is selected and assigned to the corresponding core (hence the name Min_min). Last, the newly mapped task is removed from $U$, and the process repeats until all tasks are mapped (i.e., $U$ is empty). Min_min considers all unmapped tasks on all processors during each mapping decision. Thus its algorithm complexity is $O(\chi * \gamma^2)$, where $\chi$ is the number of optional cores and $\gamma$ is the number of tasks in *Metatask*.

In [34] Min-min has been proven to have the best comprehensive performance. It is still incapable in cases where exists a big performance difference between cores, or exists a large number of short-time tasks. Min-min may encounter the problems of load unbalance and long makespan. In view of these shortcomings, here we propose an improved independent task scheduling algorithm named Diff-Min as follows:

(1) The ratio of the best execution time over the worst one of one computing unit,

$$Div(t_i) = \frac{\max_{j=1}^{|x|}(ETC(t_i, c_j))}{\min_{j=1}^{|x|}(ETC(t_i, c_j))} \quad (eq.1)$$

(2) The difference between the best and the worst execution time of each task on each computing unit:

$$Sub(t_i) = \max_{j=1}^{|x|}(ETC(t_i, c_j)) - \min_{j=1}^{|x|}(ETC(t_i, c_j)) \quad (eq.2)$$

(3) M Table：M table records the sum of all task execution times that each computing unit had allocated, and this table shows the load of each computing unit. It is a linear table of 1* x dimension, where x is the number of computing units. The core idea of the Inter-Core scheduling is to schedule the tasks using the difference in their processing performance. Dif($t_i$) and Sub($t_i$) respectively indicates the the absolute performance difference and relative performance difference of task $t_i$ processed on the cores. In another word, it shows how much benefit we can get get if task $t_i$ is properly scheduled and then accelerated. Beginning with $U=$ Metatask, we first find out the task $t_i$ with Div($t_i$)= Min(Div(t), )map the task ti to the computing unit $c_j$ that can complete all the assigned tasks and ti in the shortest time, i.e. Min(M($c_j$) + ETC(ti+$c_j$)). Once task ti is dispatched, we update the M Table by adding the execution time of ti in M($c_j$). The computing unit with minimal sum result will be the first target computing unit for next dispatch since it provides the most available time slots for future execution. Then task $t_i$ finishes, we remove it from task queue $U$ and we repeat this process until all tasks have been scheduled , i.e $U$ is empty.

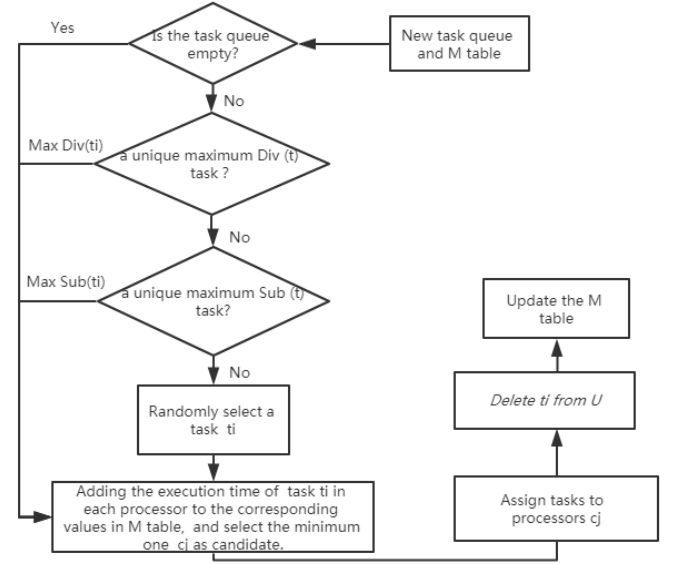

Fig 8. Execution Flow of Inter-Core Scheduling

If the task with the largest Div ($t_i$) is not unique, we give priority to the task with largest Sub ($t_i$) and dispatches this task first. If the task with the maximum Sub ($v_i$) is not unique as well, we then apply heuristics to randomly select a task from among the tasks with equal highest priority. The selection of task ti is always determined by the performance difference each optional tasks will experience with different cores. The defined performance difference can be viewed as the room for performance improvement when task is accelerated to the most extend. Thus, task ti with maximum difference deserves priority on choosing the fastest core since it can contribute the most execution time reduction. For the performance difference, we prioritize Div value over Sub value. Such design helps prevent a task from being dispatched to the computing unit with very long execution time, and thus avoiding load imbalance and task starvation problems.

The formal description of the Diff-Min algorithm is as follows:

*(1) Calculate or Estimate Div (t) and Sub (t) based on the execution time of each task on different computing units.*

*(2) Set up a task queue U and initialize it as a set of all tasks. i. e U= makespan*

*(3) Set up M table, and initialize it to all zeroes.*

*(4) Check if the task queue U is empty. If not empty, go to (5). Otherwise goes to steps (9).*

*(5) If the task ti with the maximum Div (t) value is unique, goes to (7). Else it will continue to execute.*

*(6) Compared Sub (t) value of those tasks with the maximum Div (t), if ti with maximum Sub (t) is unique, goes to (7). Otherwise, a task is randomly selected and goes to (7).*

*(7) Assign the task ti to the computing unit cj with minimal M (cj) + ETC(ti, cj).*

*(8) Remove the task from the task set U., update M table. Go to (4).*

*(9) Complete and exit.*

The time complexity of the proposed Diff-Min algorithm is ($x * \gamma$), where $x$ is the number of optional cores and $\gamma$ is the number of tasks in Metatask. Compared with Min-Min, Diff-Min can effectively reduces the algorithm complexity.

### B. Simulation Evaluation and Analysis

In this section, we use simulation to evaluate the performance of proposed Diff-Min. Since task scheduling in heterogeneous environments is a NP problem, no algorithm can achieve the optimal performance in all cases. Therefore, we will use Min-min as the baseline to compare Diff-Min.

*1) Experiment setup*

To make simulation, we need to generate three kinds of data below:
- Number of tasks. The number of tasks directly determines the size of the application and its internal complexity. The larger the number of tasks, the larger the scale of the assignment. In our simulation, we set number of tasks: $|x| \in \{10, 20, 30, 40, 50\}$;
- Number of processors: The number of processors refers to the number of heterogeneous processors that are simultaneously available for computing. In a general way, the more processors, the greater computing capacity, the shorter task completion time. In our simulation, we set number of processors: $|\gamma| \in \{3, 4, 5, 6\}$;
- Expected execution time. In heterogeneous environments, task completion time of each task on different processors is different. In our design, we use ETC table to store the expected execution time for each tasks on each processor. ETC table is a $x * \gamma$ Dimension matrix, where $x$ is the number of optional cores and $\gamma$ is the number of tasks in *Metatask*. ETC ($t_i$, $c_j$) is the expected execution time of task $t_i$ on processor $c_j$. In our simulation, we generate a random value within the range of (1,30) for each ETC ($t_i$,$c_j$) $|i< x$, $j< \gamma$ as the expected execution time.

*2) Experiment Results*

In the experiment, we generate 100 sets of task execution time data for each group with different number of tasks and processors. In Fig 9, we give the makespan comparison results of Min-min verse Diff-min. Here, all data has been normalized to the makespan under Min-min scheduling. We can clearly see that in most combination of processor number and task number, with Diff-Min the makespan can averagely be reduced by 10%. In the best case of 20 tasks on 4 processors, its acceleration can be up to 27.5%. Through further analysis, we can find that this part of efficiency mainly comes from the improvement of Min-Min algorithm in two folds: avoidance of overusing the most powerful core and more fairness on long time task. We can also find there exist a performance downgrade in the case of processor 5 and task 40. That is because in some case Diff-Min may suffer resource equalitarianism and some node will always be invalid for dispatch. However, it happens in very low frequency. With the results, we can conclude that Div-Min shows good performance in scheduling heterogeneous cores and is indeed a good solution in the scenario of autonomous driving.

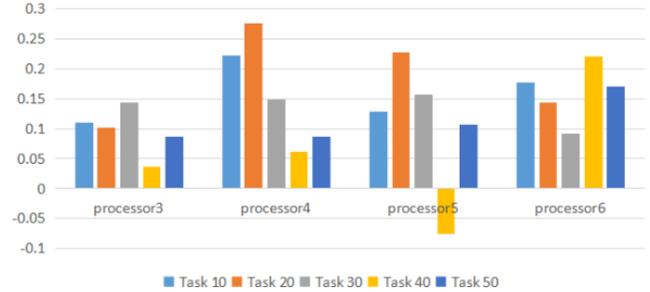

Fig 9. Comparison of Diff-Min and Min-min

### C. Inner-core Scheduling

After Inter-Core Scheduling, all computing units have been assigned with tasks, which are queued to be processed. Now the Inner-core scheduler takes over. Different from the Inter-Core scheduling, the inner-core scheduling should mine the dependency between the tasks. A task turns to be ready only after all the preceding tasks have been executed. The coupling existing in the execution, such as common coupling, data coupling, control coupling, is the root cause for the dependency. Scheduling algorithms working for such category of tasks are named dependent scheduling. Dependent scheduling problem can be described as scheduling multiple dependent tasks involved in the applications to a certain number of processors, expecting to finish the application in shortest time. In the scenario of parallel computing, the dependent scheduling should further consider how parallelism can impact the dependency execution. In literature, the table-based scheduling algorithms have attracted most attention because of its better scheduling performance and less overhead. Among them, HEFT algorithm has the best demonstration [42]. In [38], author compares HEFT algorithm, CPOP algorithm [41] and so on. The conclusion shows that no task scheduling algorithm can ensure best performance for all cases. But in a comprehensive way, HEFT can be our optimal choice.

In our inner-core scheduler design, we take a simple way to process computing on top of underlying multithreads or couples of homogeneous cores in a parallel manner. It's a two-step solution:

(1) According to the DAG task model, HEFT calculate the priority of each sub-task and rank them into several levels, and tasks on the same level do not have dependencies.

(2) After sorting, by following their sequence in the DAG graph, we schedule sub-tasks in each layer from top to bottom. The sub-tasks with the highest priority is assigned to the most appropriate cores to obtain the earliest completion time. The scheduling for next layer begins only when all the tasks in the upper layers are completed, and for the sub-tasks in the same layer we directly use Min-Min task scheduling algorithm or the proposed Diff-Min.

## V. π-OS A LIGHT-WEIGHT OPERATING SYSTEM

The second major contributions of this paper is the design and implementation of an extremely lightweight operating system, π-OS, to manage different services and facilitate their communications with almost zero overheads.

For autonomous vehicle systems, researchers are challenged to create reliable and efficient communication mediums. There are two main challenges, fast and reliable message communication, as well as lightweight software to manage the communications.

Most existing autonomous driving solutions utilize the Robot Operating System (ROS) [26], or modified versions of ROS. Specifically, ROS is a communication middleware that facilitates communications between different parts of an autonomous vehicle system. For instance, the image capture service can publish messages through ROS, and both the localization service and the obstacle detection service can subscribe to the published images to generate position and obstacle updates.

Although ROS is a popular choice for operating system in autonomous vehicle systems, in practice, it is not suitable for low-power edge computing systems due to its high communication overheads as well as the large amount of library dependencies. In our experiments, to start the basic ROS system requires roughly 50 MB of memory, even before anything service can run on it.

In addition to the high memory footprint, ROS suffers from a few communication problems: first, the current communication between ROS nodes on the same machine uses the loop-back mechanism of the network stack, which means that each data packet needs to be processed by a multi-layer software stack, which will cause unnecessary delay (20 microseconds by our measurement) with resource consumption. Second, when ROS performs a data broadcast, the underlying implementation actually uses multiple unicasts, which are multiple point-to-point transmissions. For instance, if you want to pass the data to 5 nodes, then the same data will be copied 5 copies. This causes a great waste of resources, especially the memory resources. In addition, this imposes a lot of pressure on the throughput of the communication system.

When we started this project, we did a simple hack such as to use a multicast mechanism: a point-to-multipoint network connection between the sending node and each receiving node. If a sending node transmits the same data to multiple receiving nodes at the same time, only one copy of the same data packet needs to be copied. The multicast mechanism that we implemented drastically improved data transmission efficiency and reduced the possibility of congestion in the backbone network. Figure 9 compares the performance of the original communication mechanism (gray line) with the multicast mechanism (orange). As the number of receiving nodes increases (X-axis), the data throughput of the original communication mechanism drops sharply, while the multicast mechanism The data throughput is relatively stable and has not been seriously affected.

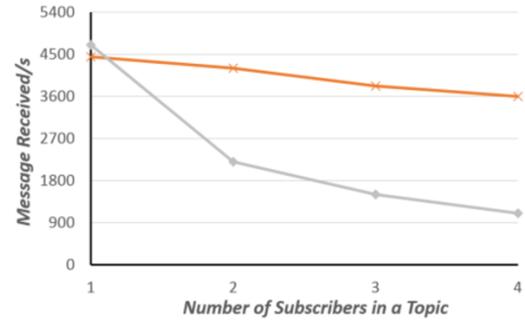

Fig 9. multicast improvement on ROS

Due to the many problems of ROS, we decided to create an extremely lightweight ROS-like system to manage all communications between the services. Our system, named π-OS, which is a crucial middle layer of π-Edge, supports multiple methods of communication: publishers, subscribers, services, and action servers. π-OS builds on top of Nanomsg, a networking library written in C that allows for easy integration of shared memory, TCP/IP, in-process messaging, and web sockets, while retaining efficiency [27]. Figure 10 shows the architecture of π-OS, which is divided into three layers. The bottom layer is the communication backbone, and we utilize Nanomsg's TCP/IP mode for inter-device communication, and shared memory for inter-process communication on the same device. On top of the backbone is the service layer, we support publish/subscribe and action server. Publish/subscribe is used to enable unidirectional data transfer between producers and any number of consumers. Action server communication is essentially a client-server service interaction pattern, as it allows servers to periodically update clients of the status of the operations they are performing by sending them feedback. On top of the services layer is the applications layer, in which we can implement different nodes to communicate with each other.

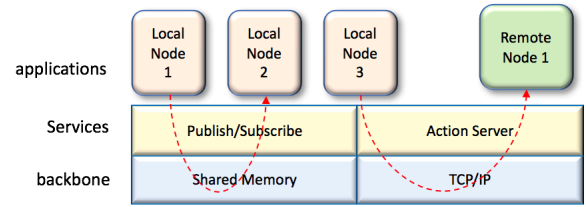

Fig 10. π-OS Architecture

Now let us examine the performance of π-OS. Based on our experiments, π-OS is able to achieve a 10 GB/s throughput, with a memory footprint of merely 10 KB for publishing and subscribing nodes, which is negligible comparing to 50 MB of ROS startup footprint. Next we performed a micro-experiment to transfer the images coming from the visual inertial sensor device to the TX1 node, and π-OS was able to stably transfer well over 60 FPS, and was able to do so with an average latency of only 0.18 ms when transferring data between the different nodes on TX1. Indeed, by our calculation, to reach the maximum transfer rate that π-OS can sustain, the visual inertial sensor would have to record images at 2,300 FPS. Next, to understand the scalability of π-OS, we studied the correlation between transfer latency and message size. Using TCP/IP loop back mechanism for inter-node communication, we confirmed that transfer latency of π-OS is directly proportional to message size. Using shared memory mechanism for inter-node communication, we confirmed that transfer latency is constant regardless of the message size.

In addition to inter-node communication on the same device, π-OS can transparently support edge client to cloud communication as well. As mentioned previously, the performance of the autonomous vehicle system can be improved by offloading some of the computation to the cloud over the network. π-OS's ultra low latency in communication makes it a good candidate for implementing the communication between the edge client and the cloud. In our edge client to a local cloud connection experiments, we found that π-OS has an average of a 50% lower latency compared to ROS, making it a suitable middleware for autonomous driving edge computing systems.

## VI. Edge Client to Cloud Offloading

The third major contributions of this paper is the design and implementation of an offloading engine that dynamically decides whether to offload a computing task, and if so, how to offload a computing task to further improve energy efficiency.

In this section, we delve into the details of the aforementioned questions, especially for the autonomous driving scenarios. Out of the three services introduced in Section II, SLAM is not a good candidate for offloading since the vehicle requires a position update every 5 ms, any delay will result in incorrect localization behavior. Beyond that, it is a good option to use offloading for the sake of energy saving. For example, for objection recognition and speech recognition, we can tolerate > 100 ms latency, making them potential candidates for offloading.

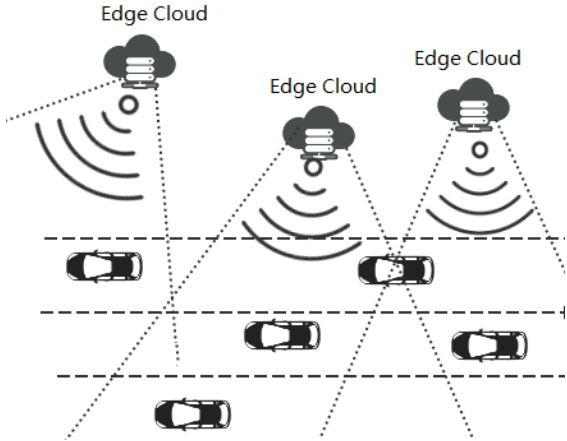

Fig 11. offloading to cloud

According to Figure 11, each edge cloud node has a fixed communication range. Within the range, the data communication cost is under the tolerance of real-time workload offloading. If out of range, the clients have to spend a large amount of time and energy for data transfer, which is not worthwhile. Moreover, in some areas, there might exist multiple edge cloud node deployments. When an edge client, like an autonomous vehicle, moves in this area, it will passes through a number of edge cloud nodes within the effective range. As a result, the optimal edge node for offloading is dynamically changing along with the vehicle's movement. Thus the core question becomes: how to select an offload service node by considering both edge client movement and edge-cloud node distributions. To address these questions, we start by defining our parameter space as follows:

**Definitions**

$m$ : the number of available edge cloud nodes;
$R_i$ : communication range of edge cloud node, $i = 1, 2, ..., m$;
$w$ : workload to be offloaded (in terms of number of instructions);
$w_i$ : remaining processing capacity in each edge cloud load, $i=1, 2, ..., m$;
$f_L$ : edge client computing speed;
$f_{off}^i$ : computing speed of edge cloud node i, $i=1,2,...,m$;
$v$ : current moving speed of the client vehicle,
$(X, Y)$ : current location of the client vehicle;
$\theta$ : current heading/orientation of movement of the client vehicle;
$D_{in}, D_{out}$ : transmission data volume as input and output;
$r_i$ : communication bandwidth between edge client and edge cloud node;
$(X_s^i, Y_s^i)$ : edge cloud node physical location, $i = 1, 2, ..., m$;
$t_c^i (i = 1, 2, ..., m)$ : the time of the vehicle staying within the communication range of edge cloud node i;

To identify the cost of making computation in the edge client versus that of offloaded to remote edge cloud node, we start by building a cost function for local computing versus offloading to edge cloud node i. Some notations first: $\eta$ is the weight to balance the need for low-power and latency. It' a user predefined value or it can be dynamically adjusted with the change of battery life. To guarantee we are making beneficial offloading, there must be a $c_{off}^i$ smaller than $c_L$.

Local computing time:
$$t_L = w / f_L \quad (eq. 3)$$

Local energy consumption:
$$e_L = \frac{(a + \beta * f_L^3) * w}{f_L} \quad (eq. 4)$$

offloading computing time:
$$t_{off}^i = w / f_{off}^i + D_{in}/r_i + D_{out}/r_i \quad (eq. 5)$$

Offloading energy consumption:
$$e_{off} = P_{in} * D_{in}/r + P_{out} * D_{out}/r \quad (eq. 6)$$

Cost function:
$$\begin{cases} c_L = \eta t_L + (1-\eta) e_L \\ c_{off}^i = \eta t_{off}^i + (1-\eta) e_{off}^i \end{cases} \quad (eq. 7)$$

$$R_i = \sqrt{(X + t_c^i * v * \sin\theta - X_c^i)^2 + (Y + t_c^i * v * \cos\theta - Y_c^i)^2}$$
(eq. 8)

Knowing the communication range of edge cloud node $R_i$ ( i= 1, 2, …, m) and its coordinate($X_c^i$, $Y_c^i$), we can get $t_c^i$ (i= 1, 2, …, m) , the time of edge client staying in the communication range of cloud node $i$. It considers the distance from the client to the edge cloud node, the speed of the moving client vehicle, and the communication coverage distance of each cloud node. For the purpose of time saving in data communication, we should guarantee that when offloading happens, the edge client is always covered within the communication range of some edge cloud node. As a result, a maximum $t_c^i$ will be the preference in solving question 2 listed above. The longer time the client vehicle stays within the communication range, the bigger time slice can be reserved for offloading computation, and the more performance and energy consumption benefits we can get through offloading in node $j$ where $t_c^j$ is the maximized in all available edge cloud nodes.

Figure 12 shows the pseudo-code for task offloading: with $t_L$, $c_L$, $e_L$, $t_{off}^i$, $e_{off}^i$, $c_{off}^i$, $t_c^i$ (i= 1, 2, …, m) , the module executes in two step: first the algorithm identifies the the optimal node for offloading by searching for a node with the maximum $t_c^i$ . Our algorithm considers both client vehicle movements and edge cloud nodes distributions, and attempts to find out the node that can connect with the moving client vehicle for the longest amount of time. Second, our algorithm estimates the performance and energy consumption benefits of offloading using the cost functions shown in equation 7. We also check if the node holds enough available computing capacity for offloading computing. If the selected node is limited in computing capacity, it will hand over the offload to the node with the second highest score.

```
for(i = 1; i ≤ M-1; i++)
    T_s = t_c^i;
    K = 0;
    for(j = 1; j ≤ M-1; j++)
        if( T_s ≤ t_c^{j+1} )
            T_s = t_c^{j+1}
            K = j + 1;
        end if
    end for
    if(T_s > t_c^i)
        switch(t_c^i, t_c^k)
        switch(w_i, w_k)
        switch(M_i, M_k)
    end if
    if( w_i > W && C_{off}^j < C_L && t_{off}^i < t_L)
        return M_i;
    end if
end for
```

Fig 12. offloading algorithm pseudo-code

## VII. π-EDGE IMPLEMENTATION ON JETSON

In this section we implement the aforementioned π-Edge architecture with the runtime layer, the π-OS, and the offloading engine, onto a Nvidia Jetson [24]. We examine the detailed performance and power consumption of such implementation and demonstrate that we could successfully support multiple autonomous driving services with only 11 W of power consumption, and thus proving the effectiveness of the proposed π-Edge system.

### A. Hardware Setup

The system consists of four parts: the sensing unit, the perception unit, and the decision unit, which are implemented on Jetson TX1, and the execution unit, which is the vehicle chassis. The vehicle chassis receives commands from the Jetson TX1, and executes the commands accordingly. A 2200 mAh battery is used to power the Jetson TX1 board.

The Jetson TX1 SoC consists of a 1024-GFLOP Maxwell GPU, a 64-bit quad-core ARM Cortex-A57, and hardware H.265 encoder/decoder. In addition, onboard components include 4GB LPDDR4, 16GB eMMC flash, 802.11ac WiFi, Bluetooth 4.0, Gigabit Ethernet, and accepts 5.5V-19.6VDC input. Peripheral interfaces consist of up to six MIPI CSI-2 cameras (on a dual ISP), 2x USB 3.0, 3x USB 2.0, PCIe gen2 x4 + x1, independent HDMI 2.0/DP 1.2 and DSI/eDP 1.4, 3x SPI, 4x I2C, 3x UART, SATA, GPIO, and others. Jetson TX1 draws as little as 1 watt of power or lower while idle, around 8-10 watts under typical CUDA load, and up to 15 watts TDP when the module is fully utilized. The four ARM A57 cores automatically scale between 102 MHz and 1.9 GHz, the memory controller between 40MHz and 1.6GHz, and the Maxwell GPU between 76 MHz and 998 MHz.

Regarding the hardware setup, a visual inertial camera module [25] is connected to the TX1 board. This module generates high-resolution stereo images at 60 FPS along with IMU updates at 200 Hz. This raw data is fed to the SLAM pipeline to produce accurate location updates, and fed to the CNN pipeline to perform object recognition. In addition, the TX1 board is connected to the underlying chassis through a serial connection. This way, after going through the sensing, perception, and decision stages, TX1 sends commands to the underlying chassis for navigation purpose. For instance, after the SLAM pipeline produces a map of the environment, the decision pipeline can instruct the vehicle to move from location A to location B, and the commands are sent through the serial interface. For speech recognition, to emulate commands, we initiate a thread to constantly perform audio playback to the speech recognition pipeline.

### B. System Architecture

Once we have made a decision on the hardware setup, the next challenge is to design a system architecture to tightly integrate these services. Figure 13 presents the architecture of the system we implement on the Jetson TX1. At the front end, we have three sensor threads to generate raw data: the camera thread generates images at a rate as high as 60 Hz, the IMU thread generates inertial updates at a rate of 200 Hz, and the microphone thread generates audio signal at a rate of 44 KHz. The image and IMU data then get into the SLAM pipeline to produce a position update at a rate of 200 Hz. Meanwhile, as the vehicle moves, the SLAM pipeline also extends the environment map. The position updates, along with the updated map, then get passed to the navigation thread to decide how the vehicle makes its next move. The image data also gets into the object recognition pipeline to extract the labels of the objects that the vehicle encounters. The labels of the objects then get fed into the reaction unit, which contains a set of rules defining the actions to take when a specific label is detected.

For instance, a rule can be that whenever a passenger gets into the vehicle, the vehicle should greet the passenger. The audio data gets through the speech recognition pipeline to extract commands, and then commands are fed to the command unit. The command unit stores a set of predefined commands, and if the incoming command matches one in the predefined command interface, the corresponding action is triggered. For instance, we implement a command "stop", whenever the word "stop" is heard and interpreted, the vehicle stops all its ongoing actions.

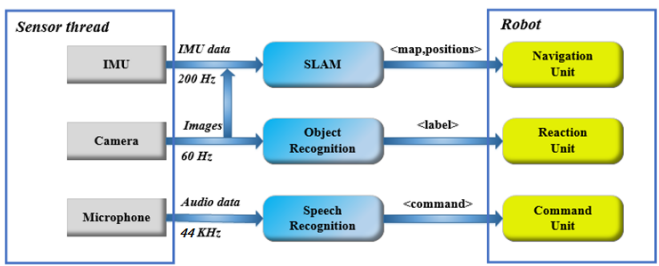

Fig 13. System Integration

This architecture provides very good separation of different tasks, with each task hosted in its own process. The key to high performance and energy efficiency is to fully utilize the underlying heterogeneous computing resources for different tasks. For instance, feature extraction operations used in the frontend of SLAM as well as CNN computations exhibit very good data parallelism, thus it would be beneficial to offload these tasks to GPU, which frees up CPU resources for other computation, or for energy efficiency. Therefore, in our implementation, the SLAM frontend is offloaded to GPU, while the SLAM backend is executed on CPU; the major part of object recognition is offloaded to GPU; the speech recognition task is executed on CPU. We will explore how this setup behaves on the Jetson TX1 SoC in the next subsections.

### C. Performance Evaluation

In this subsection we study the performance of this system. When running all the services on the system, the SLAM pipeline can process images at 10 FPS on TX1 if we use CPU only. However, once we accelerate the feature extraction stage on GPU, the SLAM pipeline can process images at 18 FPS. In our practical experience, once the SLAM pipeline is able to process images at more than 15 FPS, we have a stable localization services. As a reference, we also measured the SLAM performance on an Intel Core i5 CPU, where at its peak the SLAM pipeline processes images at 15 FPS. Therefore, with the help of GPU, the TX1 SoC can outperform a general-purpose CPU for SLAM workloads.

For the vision deep learning task using Jetson Inference engine, we can achieve 10 FPS in image recognition. This task is mostly GPU-bound. For our low-speed autonomous driving application, the vehicle travels at a fairly slow speed (at 3 m/s), where 10 FPS should satisfy our needs. For the speech recognition, we use Kaldi [10] and it is CPU-bound. We can convert an audio stream into words with 100 ms latency. In our requirement, we can tolerate 500 ms latency for such tasks. In summary, to our surprise, after we enable all these services, TX1 can still satisfy the real-time performance requirement. The main reason is that GPU performs most of the heavy lifting, especially for SLAM and vision tasks.

Next we present the system resource utilization when running these tasks. As shown in Figure 14, when running the SLAM task, it consumes about 28% CPU, 2% GPU, and 4% of system memory. The GPU is mainly used to accelerate feature extraction in this task. When running speech recognition, it consumes about 22% CPU, no GPU, and 2% system memory. For vision-based deep learning task, it consumes 24% CPU, 70% GPU, and 22% of system memory. When combining all three tasks together, the system consumes 60% CPU, 72% GPU, and 28% of system memory, still leaving enough headroom for other tasks.

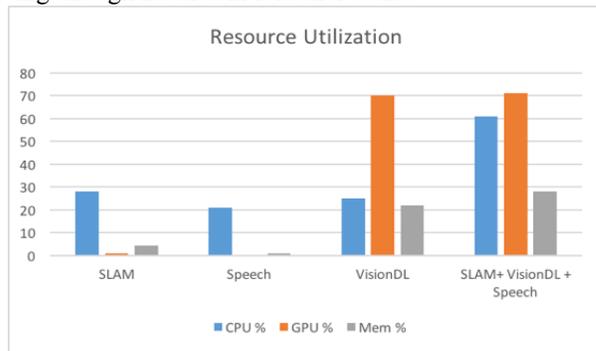

Fig 14. Resource utilization on TX1

Next we present the power consumption behavior. As shown in Figure 15, even when running all these tasks simultaneously, the TX1 module only consumes 11 W, where the GPU consumes 3.5 W, and the CPU consumes 4.2 W. In other words, with an 11 W power envelope, we can enable real-time robot localization, object recognition, and speech recognition on a TX1 SoC module.

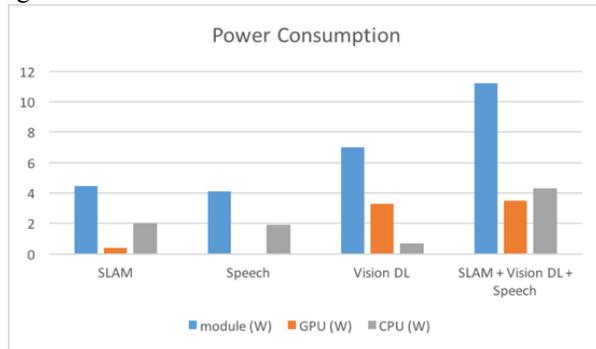

Fig 15. Power consumption on TX1

### D. Edge and Cloud Cooperation

We deploy the cloud within the local area network. After making this configuration, we have a local cloud that can perform object recognition within 100 ms and speech recognition within 200 ms, meeting the real-time requirement for the robot deployment.

Regarding resource utilization and power consumption, Figure 16 shows the resource utilization of offloading the services compared to executing locally. When offloading the tasks, we send the image or the audio file to the cloud and then wait for the results. For speech recognition, offloading consumes 5% of CPU vs. 20% CPU when executing locally. For object recognition, offloading consumes 12% CPU vs. 25% CPU and 70% GPU. When offloading object and speech recognition tasks and executing the SLAM task locally, the module consumes 5 W. Under this configuration a 2200 mAh battery can power the device for about five hours, which represents a 2.5X boost in running time.

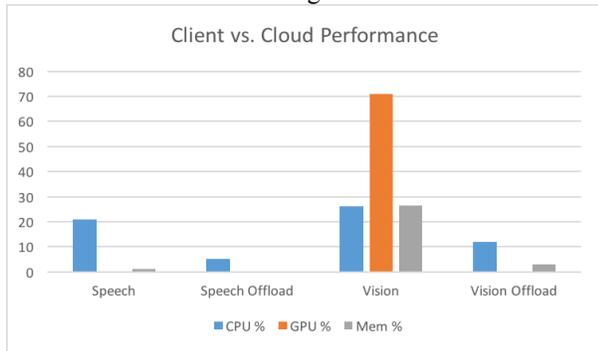

Fig 16. Client vs Cloud Performance

Based on the results from this section, we conclude that in order to optimize energy efficiency, we can deploy edge clouds to host object recognition and speech recognition tasks. Especially in a multiple-vehicle deployment, an edge cloud can be shared by the vehicles.

## VIII. RELATED WORK

Several works focus on the functionality of the autonomous driving system. In [11], Franke et al. addresses the challenges that applying autonomous driving system in complex urban traffic. They also propose an approach called Intelligent Stop & Go. Junior is the first work to introduce a full system of self-driving vehicles, which includes sensor models and deployment and software architecture design [12, 13]. Junior presents dedicated and comprehensive information about applications and software flow diagram for autonomous driving. Urmson et al develop an autonomous vehicle called Boss by using sensors including GPS, radar, camera etc [14]. Boss consists of three layers: mission planning layer, behavioral layer, and motion planning layer. Kato et al. present algorithms, libraries and datasets that are required for recognition, decision making and control [15].

There are also several works on evaluating the autonomous driving system and optimizing the performance. KITTI [16, 17] is the first benchmark suite for autonomous driving system. It comprised rich stereo image data and 2D/3D object annotated data. According to different data type, it also provided the dedicated method to generate the ground truth and calculate the evaluation metrics. CAVBench is an edge computing benchmark for Connected and autonomous vehicles, mainly focuses on the performance and power consumption of edge computing systems for autonomous vehicles. [35].

Jo et al. apply the distributed system architecture into the design of autonomous driving system [18]. And a system platform is proposed to manage the heterogeneous computing system of the distributed system. The implementation of the proposed autonomous driving system is presented in [19].

Some other works aim at building the system using heterogeneous computing platforms. Liu et al. develop an autonomous driving system architecture that can run tasks on a heterogeneous ARM mobile system on chip [20]. They divide the tasks on autonomous vehicles into three parts: sensing, perception, and decision-making. According to their experiments of performance and energy consumption on heterogeneous platform, they find that GPU is the most efficient computing units for convolutional tasks and DSP is the most efficient computing unit for feature-extraction tasks. However, the task scheduling can make the system be more complex.

In [21], an autonomous driving system based on current award-winning algorithms is implemented and they find three computational bottlenecks for CPU based systems. They compare the performance when heterogeneous computing platforms including GPUs, FPGAs, and ASICs is used to accelerate the computation. With the acceleration approach, their system can meet the performance constraints for autonomous driving system. However, more works can be done on the design of the system to promote the performance except for using hardware to accelerate the algorithms. In [22], Gao propose a safe SOC system architecture. And the security level of autonomous driving application is also discussed. However, the performance is not considered in the design of the system.

Recently, some work begins to enable edge computing in autonomous driving system. Zhang et al. propose an Open Vehicle Data Analysis Platform (OpenVDAP) for connected and autonomous vehicles [23]. OpenVDAP is a full-stack edge based platform including vehicle computing unit, an isolation-supported and security & privacy-preserved vehicle operation system, an edge-aware application library, as well as task offloading and scheduling strategy. OpenVDAP allows connected and autonomous vehicles to dynamically examine each task's status, computation cost and the optimal scheduling method so that each service could be finished in near real time with low overhead. Meanwhile, safety and security can also be a vital factor in the design of autonomous driving system.

## IX. CONCLUSIONS

Affordability is the main barrier blocking the ubiquitous adoption of autonomous driving. One of the major contributors to the high cost is the edge computing system, which can easily cost over $20,000 each. To address this problem, we built an affordable and reliable autonomous vehicle, the DragonFly pod, and we target low-speed scenarios, such as university campuses, industrial parks, and areas with limited traffic.

Within this cost structure, we had to simultaneously enable localization, perception, and speech recognition workloads on an affordable and low-power edge computing system. This was extremely challenging as we had to manage these autonomous driving services and their communications with minimal overheads, fully utilize the heterogeneous computing resources on the edge device, and offload some of the tasks to the cloud for energy efficiency.

To meet these challenges, we developed π-Edge, an edge computing framework consists of an extremely lightweight operating system, π-OS, to manage multiple autonomous driving services and their communications with almost zero overheads; a runtime layer to fully utilize the heterogeneous computing resources of low-power edge computing systems; and an edge-cloud coordinator to dynamically offload tasks to the cloud to optimize edge computing system energy consumption. As far as we know, this is the first complete edge computing system of a production autonomous vehicle.

The results were encouraging: we implemented π-Edge on a Nvidia Jetson TX1 and we demonstrated that we could successfully support vehicle localization, obstacle detection, and speech recognition services simultaneously, with only 11 W of power consumption, and hence proving the effectiveness of the proposed π-Edge system.

In the next step, we plan to extend π-Edge to support more heterogeneous edge computing architectures with more diverse computing hardware, including DSP, FPGA, and ASIC accelerators [32-34]. Besides low-speed autonomous driving, we believe π-Edge has much broader applications: by porting π-Edge to more powerful heterogeneous edge computing systems, we can deliver the computing power to L3/L4 autonomous driving; and with more affordable edge computing systems, π-Edge can be applied for delivery robots, industrial robots, *etc*.